\documentstyle[graphicx]{mn}

\newif\ifAMStwofonts

\ifoldfss
  \ifCUPmtlplainloaded \else
    \NewTextAlphabet{textbfit} {cmbxti10} {}
    \NewTextAlphabet{textbfss} {cmssbx10} {}
    \NewMathAlphabet{mathbfit} {cmbxti10} {} 
    \NewMathAlphabet{mathbfss} {cmssbx10} {} 
  \fi
  \ifAMStwofonts
    \ifCUPmtlplainloaded \else
      \NewSymbolFont{upmath} {eurm10}
      \NewSymbolFont{AMSa} {msam10}
      \NewMathSymbol{\upi}     {0}{upmath}{19}
      \NewMathSymbol{\umu}     {0}{upmath}{16}
      \NewMathSymbol{\upartial}{0}{upmath}{40}
      \NewMathSymbol{\leqslant}{3}{AMSa}{36}
      \NewMathSymbol{\geqslant}{3}{AMSa}{3E}

       \let\ge=\geqslant
    \fi
  \fi
\fi 

\ifnfssone
  \newmathalphabet{\mathit}
  \addtoversion{normal}{\mathit}{cmr}{m}{it}
  \addtoversion{bold}{\mathit}{cmr}{bx}{it}
  \newmathalphabet{\mathbfit} 
  \addtoversion{normal}{\mathbfit}{cmr}{bx}{it}
  \addtoversion{bold}{\mathbfit}{cmr}{bx}{it}
  \newmathalphabet{\mathbfss} 
  \addtoversion{normal}{\mathbfss}{cmss}{bx}{n}
  \addtoversion{bold}{\mathbfss}{cmss}{bx}{n}
  \ifAMStwofonts
    \ifCUPmtlplainloaded \else
      \UseAMStwoboldmath
      \makeatletter
      \new@mathgroup\upmath@group
      \define@mathgroup\mv@normal\upmath@group{eur}{m}{n}
      \define@mathgroup\mv@bold\upmath@group{eur}{b}{n}
      \edef\UPM{\hexnumber\upmath@group}
      \new@mathgroup\amsa@group
      \define@mathgroup\mv@normal\amsa@group{msa}{m}{n}
      \define@mathgroup\mv@bold\amsa@group{msa}{m}{n}
      \edef\AMSa{\hexnumber\amsa@group}
      \makeatother
      \mathchardef\upi="0\UPM19
      \mathchardef\umu="0\UPM16
      \mathchardef\upartial="0\UPM40
      \mathchardef\leqslant="3\AMSa36
      \mathchardef\geqslant="3\AMSa3E

       \let\ge=\geqslant
    \fi
  \fi
\fi 

\ifnfsstwo
  \DeclareMathAlphabet{\mathbfit}{OT1}{cmr}{bx}{it}
  \SetMathAlphabet\mathbfit{bold}{OT1}{cmr}{bx}{it}
  \DeclareMathAlphabet{\mathbfss}{OT1}{cmss}{bx}{n}
  \SetMathAlphabet\mathbfss{bold}{OT1}{cmss}{bx}{n}
  \ifAMStwofonts
    \ifCUPmtlplainloaded \else
      \DeclareSymbolFont{UPM}{U}{eur}{m}{n}
      \SetSymbolFont{UPM}{bold}{U}{eur}{b}{n}
      \DeclareSymbolFont{AMSa}{U}{msa}{m}{n}
      \DeclareMathSymbol{\upi}{0}{UPM}{"19}
      \DeclareMathSymbol{\umu}{0}{UPM}{"16}
      \DeclareMathSymbol{\upartial}{0}{UPM}{"40}
      \DeclareMathSymbol{\leqslant}{3}{AMSa}{"36}
      \DeclareMathSymbol{\geqslant}{3}{AMSa}{"3E}

       \let\ge=\geqslant
    \fi
  \fi
\fi 

\ifCUPmtlplainloaded \else
  \ifAMStwofonts \else 
    \def\upi{\pi}
    \def\umu{\mu}
    \def\upartial{\partial}
  \fi
\fi

\title{Statistics of dark matter haloes expected from weak lensing surveys}
\author[Guido Kruse \& Peter Schneider]
       {Guido Kruse \& Peter Schneider\\
        Max-Planck-Institut f\"ur Astrophysik, Postfach 1523, D-85740,
Garching, Germany}
\date{Accepted .
      Received ;
      in original form }

\pagerange{\pageref{firstpage}--\pageref{lastpage}}
\pubyear{1998}

\begin{document}

\maketitle

\label{firstpage}

\begin{abstract}

The distortion of the images of faint high-redshift galaxies can be
used to probe the intervening mass distribution. This weak
gravitaional lensing effect has been used recently to study the
(projected) mass distribution of several clusters at intermediate and
high redshifts. In addition, the weak lensing effect can be employed
to detect (dark) matter concentrations in the Universe, based on their
mass properties alone. Thus it is feasible to obtain a mass-selected
sample of `clusters', and thereby probe the full range of their
mass-to-light ratios. We study the expected number density of such
haloes which can be detected in ongoing and future deep wide-field
imaging surveys, using the number density of haloes as predicted by
the Press-Schechter theory, and modeling their mass profile by the
`universal' density profile found by Navarro, Frenk \& White. We find
that in all cosmological models considered, the number density of
haloes with a signal-to-noise ratio larger than 5 exceeds 10 per
square degree. With the planned MEGACAM imaging
survey of $\sim 25 \ {\rm deg}^2$, it will be easily possible to
distinguish between the most commonly discussed cosmological parameter
sets. 

\end{abstract}

\begin{keywords}
cosmology -- gravitational lensing -- clusters of galaxies -- dark
matter 
\end{keywords}

\section{Introduction}

As first discussed by Webster (1985), the tidal gravitational field of
clusters of galaxies distorts the images of background galaxies in a
characteristic way. After the first extreme cases of distortions in
the form of giant luminous arcs were discovered (Soucail et al.\ 1987; see
Fort \& Mellier 1994 for a review), much weaker coherent distortions
of images were found (Fort et al.\ 1988; Tyson, Valdes \& Wenk
1990). These distortions can be used to reconstruct the projected mass
distribution of galaxy clusters in a non-parametric way (Kaiser \& Squires
1993).

In addition, as pointed out in Schneider (1996; hereafter S96), the
search for coherent image alignments can be used to search for (dark)
mass concentrations. Generalizing the aperture densitometry of Kaiser
(1995; see also Kaiser et al.\ 1994), it was shown in S96 that halos
with characteristic velocity dispersions of $\ge 600\;$km/s can be
significantly detected on deep high-quality optical images, such as
can be obtained with a 4-metre class telescope at the best
sites. Indeed, there are first reports of detections of mass
concentrations selected by this weak lensing technique
(Luppino \& Kaiser 1997; T.\ Erben, private communication) which
coincide with a concentration of galaxies; they are most likely genuine
clusters. Seitz et al.\ (1998) have provided a thorough lensing
analysis of the cluster MS1512+36 which acts as a strong gravitational
telescope on the $z=2.72$ galaxy cB58 detected by Yee et
al. (1996). Whereas this cluster appears to have a velocity dispersion
of order 600 km/s, it nevertheless shows up with very high significance
in the weak lensing analysis, using only 33 (background) galaxies and
excluding the strong lensing features, thus observationally verifying
the estimate of S96. 

The detection of mass concentrations by weak lensing techniques
therefore offers the opportunity to define a mass-selected sample of
haloes. In contrast to the usual selection procedures, based on
emitted light (in the optical or X-ray waveband), the resulting sample
would be `mass-limited', rather than of flux limited. Such a sample would
therefore be extremely useful for cosmological purposes, since it can
directly be compared to theoretical predictions, e.g., derived from
N-body simulations.  
In contrast, the comparison of optically-selected
cluster samples with cosmological predictions involves
assumptions about the relation between mass and light, and the
mass-to-light ratio may vary strongly between individual
clusters. Given that the evolution of clusters with redshift 
is among the strongest tests for distinguishing between different
cosmogonies (see, e.g., White, Efstathiou \& Frenk 1993; Eke, Cole \&
Frenk 1996; Bartelmann et al. 1998;
Borgani et al.\ 1998, and references therein), their
mass-based detection would indeed be of great interest. A
mass-selected sample of haloes may lead to the detection of clusters
with very faint emission which could be missed by other selection
criteria. 

The basic method discussed in S96 is to use the aperture mass $M_{\rm
ap}(\theta)$ technique (Kaiser 1995; Squires \& Kaiser 1996) on
deep wide-field images. The aperture mass is the projected density
field of the mass inhomogeneities between us and the population of
faint high-redshift galaxies, weighted by a redshift-dependent term
and filtered through a function of zero net weight (e.g., a
Mexican hat). The advantage of this measure is that it can be
expressed directly in terms of the shear, for which the observed image
ellipticities provide an unbiased estimate. Thus, an estimate for the
aperture mass can be expressed directly in terms of observables, with
well defined signal-to-noise ratio. Hence, a (dark) matter
concentration would be `seen' as a high S/N peak in the aperture mass
map. 

In this paper, we investigate the statistics of such peaks in various
cosmological models. The number density of haloes is calculated using
the Press-Schechter (1974) formalism, and their density profile is
approximated by the universal halo profile found by Navarro, Frenk \&
White (1996, 1997; hereafter NFW). In Sect.\ 2 we summarize our
method, and estimate signal-to-noise statistics in Sect.\ 3. The
number of haloes of given $M_{\rm ap}(\theta)$, as a function of
filter scale $\theta$, and source and lens redshift, is derived in
Sect.\ 4. We discuss the degree to which observations can
be used to distinguish between these various cosmologies in Sect.\
5, and present our conclusions in Sect.\ 6.
	
\section{Formalism}

Following S96, we define
the spatially filtered mass inside a circular aperture of angular
radius $\theta$,
\begin{equation}
M_{\rm ap} (\theta):=\int d^{2} \vartheta \ \kappa(
\mbox{\boldmath$\vartheta$}) 
\ U(\vert \mbox{\boldmath$\vartheta$} \vert), 
\label{aperture}
\end{equation}
where
the continuous weight function $U(\vartheta)$ vanishes for
$\vartheta>\theta$.
If $U(\vartheta)$ is a compensated filter function,
\begin{equation}
\int_0^{\theta} d \vartheta \ \vartheta \ U(\vartheta)=0,
\end{equation} 
one can express $M_{\rm ap}$ in terms of the tangential shear inside the
circle
\begin{equation}
M_{\rm ap} (\theta) =\int d^{2} \vartheta \ 
\gamma_{\rm t}(\mbox{\boldmath$\vartheta$}) 
\ Q(\vert \mbox{\boldmath$\vartheta$} \vert),
\label{mapsh}
\end{equation}
where
\begin{equation}
\gamma_{\rm t}(\mbox{\boldmath$\vartheta$}) =
-{\rm Re}(\gamma(\mbox{\boldmath$\vartheta$})
e^{-2i \phi})
\label{tanshear}
\end{equation}
is the tangential component of the shear at position $\mbox{\boldmath$
\vartheta$}=(\vartheta \ \mbox{cos} \ \phi,\vartheta \ \mbox{sin} \ \phi)$
, and the function $Q$ is related to $U$ by
\begin{equation}
Q(\vartheta) = \frac{2}{\vartheta^2} \ \int_0^{\vartheta}
d \vartheta^{\prime} \ \vartheta^{\prime} \ U(\vartheta^
{\prime})
\ - U(\vartheta) .
\end{equation}
We use a filter function from the familiy given in Schneider et
al. (1998), specifically we choose the one with $l=1$. 
Then writing $U(\vartheta)=u(\vartheta/
\theta)/\theta^2$, and $Q(\vartheta)=
q(\vartheta/\theta)/\theta^2$, 
\begin{equation}	
u(x)=\frac{9}{\pi} (1-x^2)\left( \frac{1}{3}-x^2 \right), 
\end{equation}
and
\begin{equation}	
q(x)=\frac{6}{\pi} x^2(1-x^2),
\end{equation}
with $u(x)=0$ and $q(x)=0$ for $x>1$.
We will describe the mass density of dark matter haloes with the
universal density profile introduced by NFW,
\begin{equation}
\rho (r)= \frac{3 H_0^2}{8 \pi G} \ (1+z)^3 \ \frac{\Omega_{\rm d}}{\Omega
(z)} \ \frac{\delta_c}{r/r_{\rm s} (1+r/r_{\rm s})^2},  
\end{equation}
with
\begin{equation}
\Omega(z)=\frac{\Omega_{\rm d}}{a+\Omega_{\rm d} (1-a)+\Omega_{\rm v} 
(a^3-a)}, \ a=\frac{1}{1+z}.  
\end{equation}
$\Omega_{\rm d}$ and $\Omega_{\rm v}$ denote the present day density
parameters in dust and in vacuum energy respectively.
Haloes identified at redshift $z$ with mass $M$ are described by the
characteristic density contrast $\delta_{\rm c}$ and the scaling radius 
$r_{\rm s}=r_{200}/c$ where $c$ 
is the concentration parameter (which is a function of
$\delta_{\rm c}$), and $r_{200}$ is the virial radius 
defined such that a sphere with radius $r_{200}$ of mean interior
density $200 \ \rho_{\rm crit}$ contains the halo mass $M_{200}$.
We compute the parameters which specify
the NFW profile according to the description in NFW using the fitting
formulae given there.
 
The surface mass density of the NFW-profile is given by 
(see Bartelmann 1996)	
\begin{equation}
\Sigma (\vartheta)=  \frac{3 H_0^2}{4 \pi G} \ (1+z)^3 \ 
\frac{\Omega_{\rm d}}{\Omega
(z)} \ r_{\rm s} \ \delta_{\rm c} \
f \left( \frac{\vartheta}{\theta_{\rm s}} \right)
\end{equation}
with
\begin{eqnarray}
f(x)&=&\frac{1}{x^2-1} \ \times \nonumber \\
&\biggl\{&^{\displaystyle{
1-\frac{2}{\sqrt{1-x^2}} \ \mbox{arctanh} \ 
\sqrt{\frac{1-x}{1+x}}, \ \mbox{for} \ x < 1}}
_{\displaystyle{1-\frac{2}{\sqrt{x^2-1}} \ \mbox{arctan} \ 
\sqrt{\frac{x-1}{1+x}}, \ \mbox{for} \ x > 1}},
\end{eqnarray}
and $\theta_{\rm s}= r_{\rm s}/D_{\rm d}$. $D_{\rm d}$ is the 
angular diameter distance to the lens.
Introducing the critical surface density
\begin{equation}
\Sigma_{\rm cr}= \frac{c^2}{4 \pi G} \ \frac{D_{\rm s}}
{D_{\rm d} D_{\rm ds}} 
\label{geom}
\end{equation}
with $D_{\rm s}$ and $D_{\rm ds}$ being the angular
diameter distances
to the source and from the lens to the source,
we define the dimensionless surface mass density (convergence)
which is a function of source redshift
\begin{equation}
\kappa (\vartheta,z_{\rm d},z_{\rm s}) = \frac{\Sigma (\vartheta)}
{\Sigma_{\rm cr}} =
\kappa_0 \ f \left( \frac{\vartheta}{\theta_{\rm s}} \right),
\end{equation}
with
\begin{equation}
\kappa_0=3 \ (1+z)^3 \ \frac{\Omega_{\rm d}}{\Omega(z)}\ r_{\rm s} \
\frac{H_0^2}{c^2} \ \delta_c \ \frac{D_{\rm d} D_{\rm ds}}{D_{\rm s}}. 
\end{equation}

The second important quantity for lensing effects is the complex shear
defined by
\begin{equation}
\gamma = \gamma_1+i \gamma_2, \ \gamma_1=\frac{1}{2} (\psi_{11}-
\psi_{22}), \ \gamma_2=\psi_{12},
\end{equation}
where $\psi$ is given by the two-dimensional Poisson
equation
\begin{equation}
\nabla^2 \psi = 2 \kappa.
\end{equation}
In the case of an axi-symmetric density profile, the magnitude of
the shear is given by
\begin{equation}
\vert \gamma(\vartheta) \vert = \left\vert \frac{m(\vartheta)}{\vartheta^2} 
- \kappa(\vartheta) \right\vert,
\end{equation}
where
\begin{equation}
m(\vartheta)=2 \int_0^{\vartheta} \ d \vartheta' 
\ \vartheta' \kappa(\vartheta').
\end{equation}
We obtain 
\begin{equation}
\vert \gamma \vert (\vartheta,z_{\rm d},z_{\rm s})= \kappa_0 \ 
g \left( \frac{\vartheta}{\theta_{\rm s}} \right),
\end{equation}
with
\begin{equation}
g(x)= 
\frac{2}{x^2} 
{\rm ln} \frac{x}{2}-
\frac{1}{x^2-1}+ \frac{6x^2-4}{x^2(x^2-1)^{1.5}}\mbox{arctan} 
\sqrt{\frac{x-1}{x+1}}, 
\end{equation}
$\mbox{for} \ x>1$, and 
\begin{eqnarray}
&&g(x)=\nonumber \\
&&\frac{2}{x^2} {\rm ln} \frac{x}{2}-\frac{1}{x^2-1}+
\frac{4-6x^2}{x^2(1-x^2)^{1.5}}\mbox{arctanh} \sqrt{\frac{
1-x}{x+1}},
\end{eqnarray}
$\mbox{for} \ x<1$.
According to eq.(4) the tangential shear is
\begin{equation}
\gamma_{\rm t} (\vartheta)= 
\frac{m(\vartheta)}{\vartheta^2} - \kappa(\vartheta).
\end{equation}
We assume a normalized source redshift distribution of the form
\begin{equation}	
p_z (z)= \frac{\beta}{z_0^3 \ \Gamma \left(\frac{3}{\beta} \right)}
\ z^2 \ \mbox{exp}(-[z/z_{0}]^{\beta}),
\label{sources}
\end{equation}	 
(see Brainerd et al. 1996).
The mean redshift of this distribution is proportional to 
$z_{0}$ and depends on the parameter $\beta$ which describes
how quickly the distribution falls off towards higher redshifts.
We will use the values $\beta=1.5$ and $z_0=1$. For these values
the mean redshift $\langle z \rangle$ is given by
$\langle z \rangle =1.505 \ z_{0}$.
With the distribution (\ref{sources}) we define 
a source distance-averaged surface density and shear	
\begin{equation}	
\kappa (\vartheta,z_{\rm d}) = \int dz_{\rm s} \ p_{z} 
(z_{\rm s}) \ \kappa (\vartheta,z_{\rm d},z_{\rm s}), 
\end{equation}
\begin{equation}	
\gamma_{\rm t} (\vartheta,z_{\rm d}) = \int dz_{\rm s} 
\ p_{z} (z_{\rm s}) \ 
\gamma_{\rm t} (\vartheta,z_{\rm d},z_{\rm s}).      	
\end{equation}
We emphasise here that the aperture mass $M_{\rm ap}$ in this form
depends on three parameters: the lens mass $M$, the lens redshift
$z_{\rm d}$ and the aperture radius $\theta$. The mass and redshift dependence
comes from the characteristic density $\delta_{\rm c}$, 
the scaling radius $r_{\rm s}$ and $D_{\rm d}, D_{\rm ds}$.
Furthermore, $M_{\rm ap}$ depends on cosmology
through the angular diameter distances, $\delta_{\rm c}$ and $r_{\rm s}$.

\section{Signal-To-Noise Ratio Statistics}

S96 introduced a signal-to-noise ratio for the $M_{\rm ap}$-
statistics.
An discretised estimator for (\ref{mapsh}) is given by
\begin{equation}
M_{\rm ap} = \frac{1}{n} \ \sum_i \ \epsilon_{{\rm t}} 
(\mbox{\boldmath$\vartheta_i$}) \   
Q (|\mbox{\boldmath$\vartheta_i$}|),
\label{est}
\end{equation}
where $n$ is the number density of galaxy images and $\epsilon_{\rm t}$
is the tangential component of the ellipticity of a galaxy at position
$\mbox{\boldmath$\vartheta_i$}$, defined in analogy to (\ref{tanshear}).
The dispersion $\sigma_{\rm d}$ of $M_{\rm ap}$
in the absence of lensing can be calculated by
squaring (\ref{est}) and taking the expectation value, which leads to
\begin{equation}
\sigma_{\rm d}^2 = \frac{\sigma_{\epsilon}^2}{2 n^2} \ 
\sum_i Q^2  (\vartheta_i),
\label{sigmax}
\end{equation}
where we used that the ellipticities of different 
images are not correlated and the dispersion of the observed
ellipticity equals the intrinsic ellipticity distribution
$\sigma_{\epsilon}$.
We take as reference values $\sigma_{\epsilon}=0.2$ and $n=30 \  
\mbox{arcmin}^{-2}$ (see S96). 
The expectation value of $M_{\rm ap}$ is
\begin{equation}
\langle M_{\rm ap} \rangle_{\rm d} =  \frac{1}{n} \ \sum_i \
\gamma_{\rm t}  (\vartheta_i)  \ Q (\vartheta_i),
\end{equation}
because the ellipticity is an unbiased estimate of the local
shear in the case of weak lensing.
Averaging (\ref{sigmax}) over the probability distribution
for the spatial distribution of galaxies (see S96) one finds 
\begin{equation}
\sigma_{\rm c}^2 (\theta) =  \frac{\pi \sigma_{\epsilon}^2}{n}\ 
\int^{\theta}_{0}  d \vartheta \ \vartheta \ Q^2 (\vartheta)
= 0.2 \ \frac{\sigma_{\epsilon}^2}{n} \ \frac{1}{\theta^2}.
\label{sigma}
\end{equation}
For reference, we rewrite (\ref{sigma}) in useful units as 
\begin{eqnarray}
&&\sigma_{\rm c} (\theta) =\nonumber\\
&&0.016 \left(\frac{n}{30 \ \mbox{arcmin}^{-2}}\right)^{-1/2} 
\left(\frac{\sigma_
{\epsilon}}{0.2} \right) \left(\frac{\theta}{1 \ \mbox{arcmin}} \right)^{-1}.
\end{eqnarray}	
We define the signal-to-noise ratio as 
\begin{equation}
S_{\rm c} (\theta) = \frac{M_{\rm ap} (\theta)}{\sigma_{\rm c}
(\theta)}.
\label{sig1}
\end{equation}	 
Note that $\sigma_{\rm c}$ depends only on the filter scale $\theta$
and the intrinsic properties of the source galaxies.

\section{Number of haloes}

\begin{figure}
\center{\includegraphics[
        width=\columnwidth,
       draft=false,
        ]{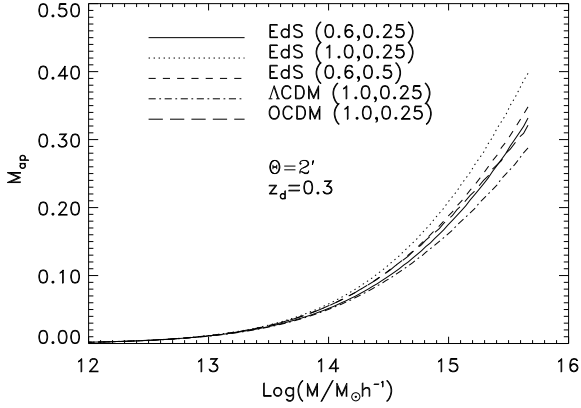}}
\caption{The aperture mass $M_{\rm ap}$, as defined in
(\ref{aperture}), computed for five different cosmologies as
indicated by the line types. The numbers in parentheses are
the normalization $\sigma_8$ and the shape parameter $\Gamma$.
The lens redshift is $z_{\rm d}=0.3$ and the aperture radius $\theta=2
\ \mbox{arcmin}$. 
\label{map}}
\end{figure}

We assume dark matter haloes 
are distributed according to the Press-Schechter (1974) theory .
In this formalism an analytical expression for the comoving
number density of 
nonlinear objects is derived on the basis of the spherical 
collapse theory assuming the initial density contrast to be a gaussian random
field. The mass fraction in collapsed objects in the mass range
$dM$ about $M$ is given by
\begin{eqnarray}	
&&f(M,z) \ dM = \nonumber\\
&&\sqrt{\frac{2}{\pi}} \ {\delta_{\rm crit} (z) \over \
\sigma^{2} (M)} \ \left \vert \frac{d \sigma(M)}{d M} \right \vert \
\mbox{exp} \left(- \frac{\delta_{\rm crit}^{2} (z)}{2 \sigma^{2}(M)}
\right) \ dM.
\label{fm}
\end{eqnarray}
The redshift
dependence of this function is given by the critical density treshold
$\delta_{\rm crit}(z)$ for spherical collaps which depends on the linear
growth factor $D_{+} (z)$ (Lacey $\&$ Cole 1993).
$\sigma(M)$ is the present linear theory rms density fluctuation
computed using a top hat filter 
and a CDM power spectrum 
(Bardeen et al.  1986) with shape parameter $\Gamma$
and normalization $\sigma_8$.
We use the fitting formulae given in NFW to compute $\sigma(M)$ 
and $\delta_{\rm crit}(z)$.	
If we multiply eq.(\ref{fm}) with $dV_{\rm p} (1+z)^{3} \bar \rho/M$ 
,where $\bar \rho$ is the mean mass density today, we get the 
number of objects in the proper volume $dV_{\rm p}$  
with mass in the intervall $dM$ 
\begin{equation}	
N_{\rm halo}(M,z) \ dM \ dV_{\rm p}= (1+z)^{3} \ \frac{\bar \rho}{M} 
\ f(M,z) \ dM \ dV_{\rm p}.
\end{equation}
For fixed values of the lens redshift $z_{\rm d}$ and the aperture
radius $\theta$
the aperture mass $M_{\rm ap}$ is a monotonically 
increasing function of the halo mass $M$
(see Figure \ref{map}). 
This function can be inverted for a given value of $M_{\rm ap}$.
We write $M_{\rm t}=M_{\rm t}(M_{\rm ap},z_{\rm
d},\theta)$ for the mass obtained by inversion.
The number of haloes in a given proper volume with mass greater 
than $M_{\rm t}(M_{\rm ap}^0,z_{\rm d},\theta)$,
and thus an aperture mass larger than $M_{\rm ap}^0$, is given by
\begin{equation}
N (>M_{\rm ap}^{0}, \theta) = \ \int \ d V_{\rm p} \ 
G(z_{\rm d},M_{\rm ap}^{0}, \theta),
\end{equation}
with 
\begin{eqnarray}
&&G(z_{\rm d},M_{\rm ap}^{0}, \theta) =\nonumber\\
&&\int dM \ N_{\rm halo}(M,z_{\rm d}) \ 
H( M_{\rm ap} (M,z_{\rm d},\theta)-M_{\rm ap}^{0}),
\end{eqnarray}
where $H(x-y)$ is the Heaviside step function.
The integral is non-zero only for $M>M_{\rm t}$.
Hence by introducing spherical polar coordinates
\begin{equation}	     
d V_{\rm p} = D_{\rm d}^2 (z_{\rm d}) \ d D_{\rm p} (z_{\rm d}), \\ 
d D_{\rm p} (z_{\rm d}) = \frac{d z_{\rm d}}{E(z_{\rm d}) (1+z_{\rm d})}, 
\end{equation}
and
\begin{equation}	
E(z_{\rm d})= \sqrt{\Omega_{\rm d} (1+z_{\rm d})^3+ (1-\Omega_{\rm d}
-\Omega_{\rm v})
(1+z_{\rm d})^2+
\Omega_{\rm v}}
\end{equation}
we obtain
\begin{equation}
N (>M_{\rm ap},\theta) = \ \int \ d z_{\rm d} \ 
\frac{(1+z_{\rm d})^2}{E(z_{\rm d})} \ D_{\rm d}^2 (z_{\rm d}) \ \tilde
G(z_{\rm d},M_{\rm ap}, \theta),
\label{number}
\end{equation}
with
\begin{equation}
\tilde G(z_{\rm d},M_{\rm ap}, \theta)= 
\int_{M_{\rm t}(M_{\rm ap},z_{\rm d},\theta)}^{\infty} 
\ dM \ n(M,z_{\rm d}),
\label{GZ} 
\end{equation}
where $n(M,z_{\rm d})= \frac{\bar \rho}{M} \ f(M,z_{\rm d})$.
$N (>M_{\rm ap},\theta)$ is the number of haloes per steradian
with aperture mass larger than $M_{\rm ap}$. Since the aperture
mass is determined by the tangential shear the number of haloes 
$N (>M_{\rm ap},\theta)$ is an observable. 

\section{Results}

In this section we use the observable $N (>M_{\rm ap},\theta)$ to
constrain various cosmological models. 
We perform our calculations for
the same five cosmological models as in Figure \ref{map}.  
For three of them, the power spectrum is approximately cluster
normalized, which corresponds to $\sigma_8 \approx 0.6$ for an
Einstein-de Sitter universe (EdS, $\Omega_{\rm d} = 1$, 
$\Omega_{\rm v} = 0$) and $\sigma_8 = 1$ for both an open universe
(OCDM, $\Omega_{\rm d} = 0.3$, $\Omega_{\rm v} = 0$) and a spatially
flat universe with cosmological constant ($\Lambda$CDM, $\Omega_{\rm d}
= 0.3$, $\Omega_{\rm v} = 0.7$). For these models we use 
the shape parameter $\Gamma = 0.25$ which
yields the best fit to the observed two-point correlation function
of galaxies (Efstathiou 1996). The remaining two 
EdS models
have higher normalization ($\sigma_8 = 1$, approximately corresponding
to the COBE normalization) or a different shape parameter ($\Gamma =
0.5$).
 
\begin{figure}
\center{\includegraphics[
        width=\columnwidth,
       draft=false,
        ]{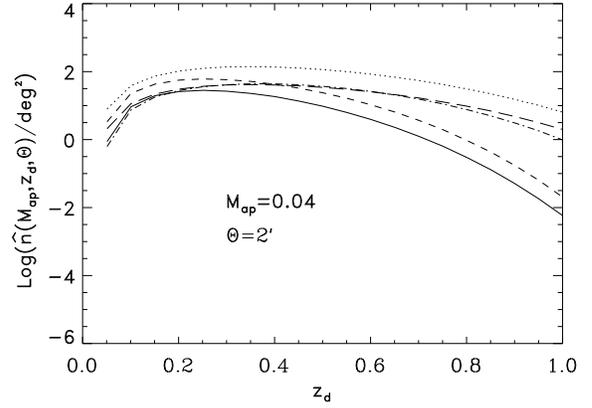}}
\caption{The number of haloes per square degree and unit redshift
interval with aperture mass greater than 0.04,
as defined in (\ref{nd}), as a function of lens redshift for
the same cosmological models as indicated in Figure \ref{map}.
The filter scale is $\theta=2$ arcmin. 
\label{ndachx}}
\end{figure}

\begin{figure}
\center{\includegraphics[
        width=\columnwidth,
       draft=false,
        ]{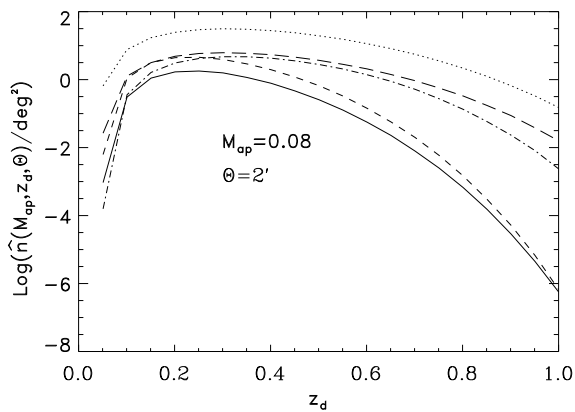}}
\caption{Same as Fig.\ \ref{ndachx}, for $M_{\rm ap}=0.08$.
The aperture mass is a monotonically increasing function of
the halo mass (see Figure \ref{map}). Furthermore, rich clusters
evolve more than clusters with low mass. Therefore, compared to
Figure \ref{ndachx}, 
the number density of haloes is smaller and we observe
a stronger evolution in the various cosmologies.
\label{ndach1}}
\end{figure}

In Figures \ref{ndachx} and \ref{ndach1} we plot the function
\begin{equation}
\hat n(M_{\rm ap},z_{\rm d},\theta) = \frac{(1+z_{\rm d})^2}
{E(z_{\rm d})} \ D_{\rm d}^2 (z_{\rm d}) \
\tilde G(z_{\rm d},M_{\rm ap}, \theta), 
\label{nd}
\end{equation}
which
is the number of haloes per unit solid angle and unit redshift interval
with masses $M>M_{\rm t}(M_{\rm ap},z_{\rm d}, \theta)$
for a filter scale of $\theta=2'$ and the aperture masses
$M_{\rm ap}=0.04$ and $M_{\rm ap}=0.08$ 
[see (\ref{GZ}) for definition of $\tilde G(z_{\rm d},M_{\rm ap},
\theta)$].
\begin{figure}
\center{\includegraphics[
        width=\columnwidth,
       draft=false,
        ]{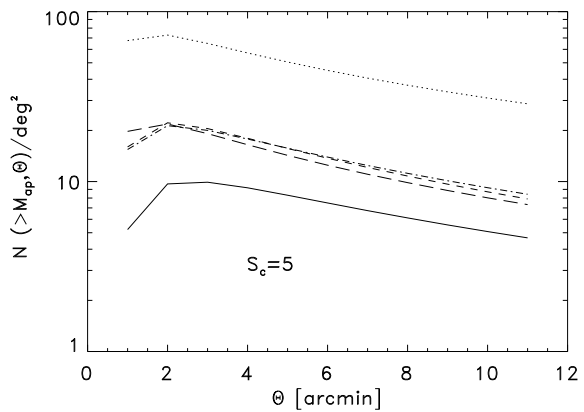}}
\caption{The number of haloes per square degree with aperture mass 
greater than 
$M_{\rm ap}=5 \ \sigma_{\rm c} (\theta)$, as defined in 
(\ref{number}), as a function of the filter scale for
the same cosmological models as indicated in Figure \ref{map}.
For $\theta=2'$ we obtain a maximum number of haloes for all
cosmologies at a fixed signal-to-noise ratio $S_{\rm c}=5$.
\label{snoise}}
\end{figure}
\begin{figure}
\center{\includegraphics[
        width=\columnwidth,
       draft=false,
        ]{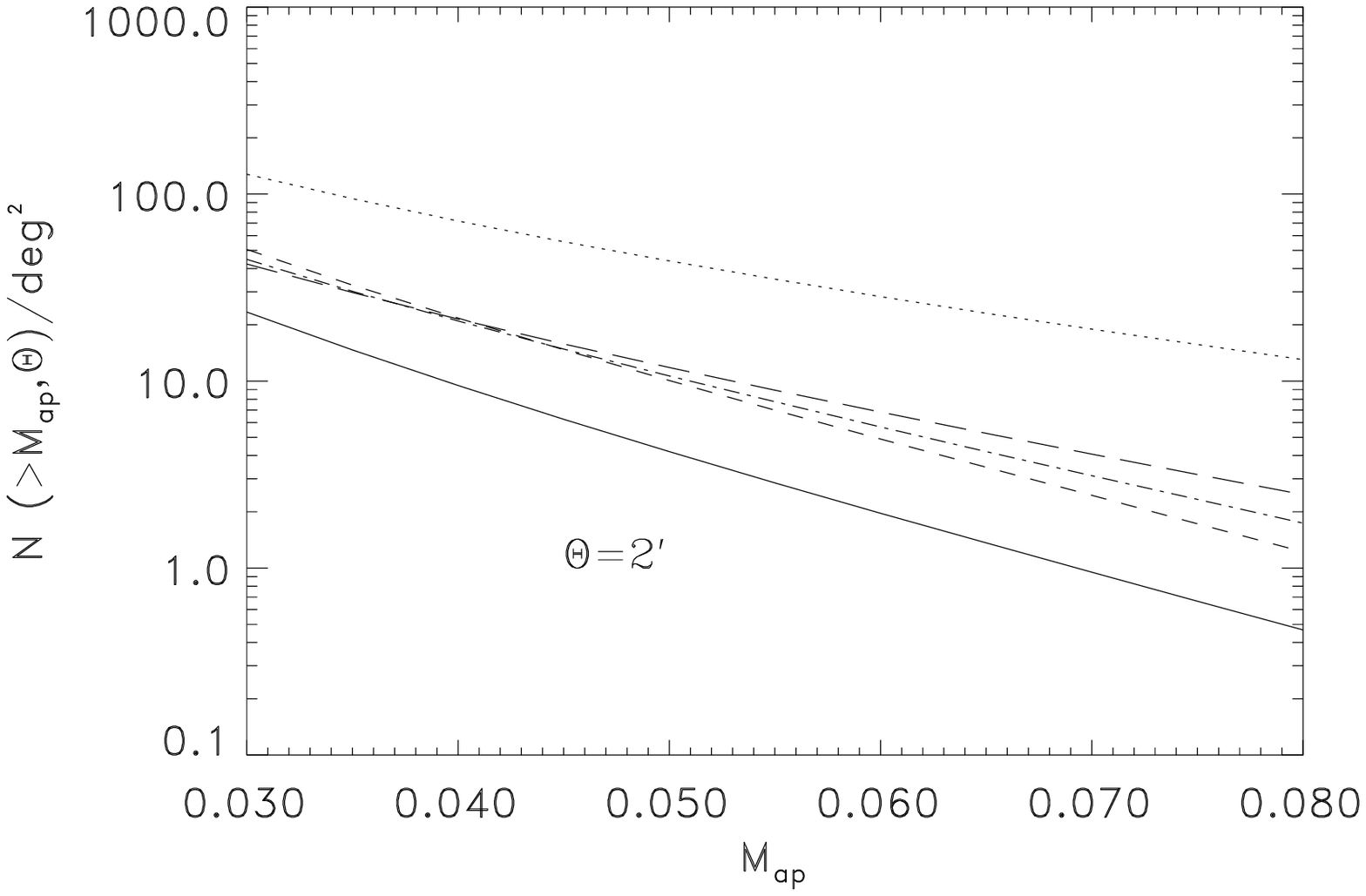}}
\caption{The number of haloes per square degree with aperture mass
greater than $M_{\rm ap}$, as defined in
(\ref{number}), as a function of $M_{\rm ap}$
for the same
cosmological models as indicated in Figure \ref{map}. 
The filter scale is $\theta=2$ arcmin.
\label{numbermap}}
\end{figure}
\begin{figure}
\center{\includegraphics[
        width=\columnwidth,
       draft=false,
        ]{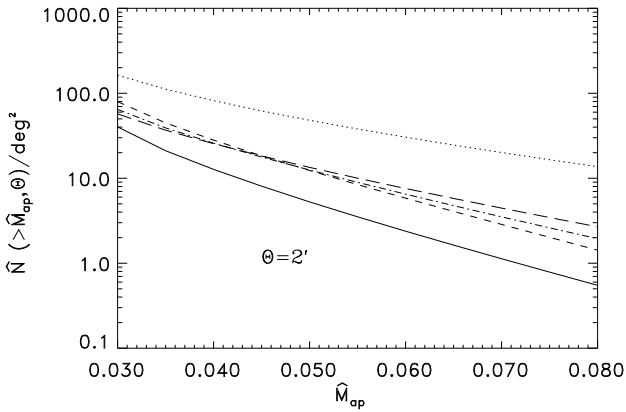}}
\caption{The convolution of the number of haloes per square degree 
with aperture mass greater than $M_{\rm ap}$ with the distribution
(\ref{prob}), as defined in
(\ref{numberhat}), as a function of $M_{\rm ap}$
for the same
cosmological models as indicated in Figure \ref{map}. 
The filter scale is $\theta=2$ arcmin. Compared to Figure
\ref{numbermap} the number of haloes is shifted to higher values.
\label{numbermapcon}}
\end{figure}
\begin{table*}
 \centering
 \begin{minipage}{140mm}
  \caption{The number of haloes per square degree with aperture mass
           greater than $M_{\rm ap}=0.04$ and $M_{\rm ap}=0.08$
           , as defined in 
           (\ref{number}), for
           the filter scale $\theta=2$ arcmin. 
           The redshift interval in brackets denote the integration
           range in (\ref{number}). The number of haloes is 
           computed for five cosmological models.
           }
  \begin{tabular}{@{}lrrrrrlrlr@{}}
          &  &  &  & \\
          & EdS(0.6,0.25) & EdS(1,0.25) & EdS(0.6,0.5) &  
     OCDM(1,0.25) & $\Lambda$CDM(1,0.25) \\
        &  &  &  &  &  & & & \\
        &  &  &  &  &  & & & \\
$N(>0.04,2')$ & 9.42 & 71.66 & 21.66 & 21.44 & 20.92 \\
$z_{\rm d} \in [0,1]$ \\  
& \\
$N(>0.08,2')$ & 0.47 & 13.00 & 1.23 & 2.46 & 1.74 \\ 
$z_{\rm d} \in [0,1]$ \\
& \\
$N(>0.04,2')$ & 6.18 & 30.66 & 13.54 & 9.07 & 9.02 \\
$z_{\rm d} \in [0.15,0.4]$ \\
& \\
$N(>0.04,2')$ & 2.34 & 37.09 & 5.89 & 11.23 & 11.14 \\
$z_{\rm d} \in [0.4,1]$ \\
& \\
$N(>0.08,2')$ & 0.36 & 6.94 & 0.93 & 1.37 & 0.99 \\
$z_{\rm d} \in [0.15,0.4]$ \\
& \\
$N(>0.08,2')$ & 0.06 & 5.29 & 0.17 & 0.97 & 0.71 \\
$z_{\rm d} \in [0.4,1]$    \\
\label{table}
\end{tabular}
\end{minipage}
\end{table*}
\begin{figure}
\center{\includegraphics[
        width=\columnwidth,
       draft=false,
        ]{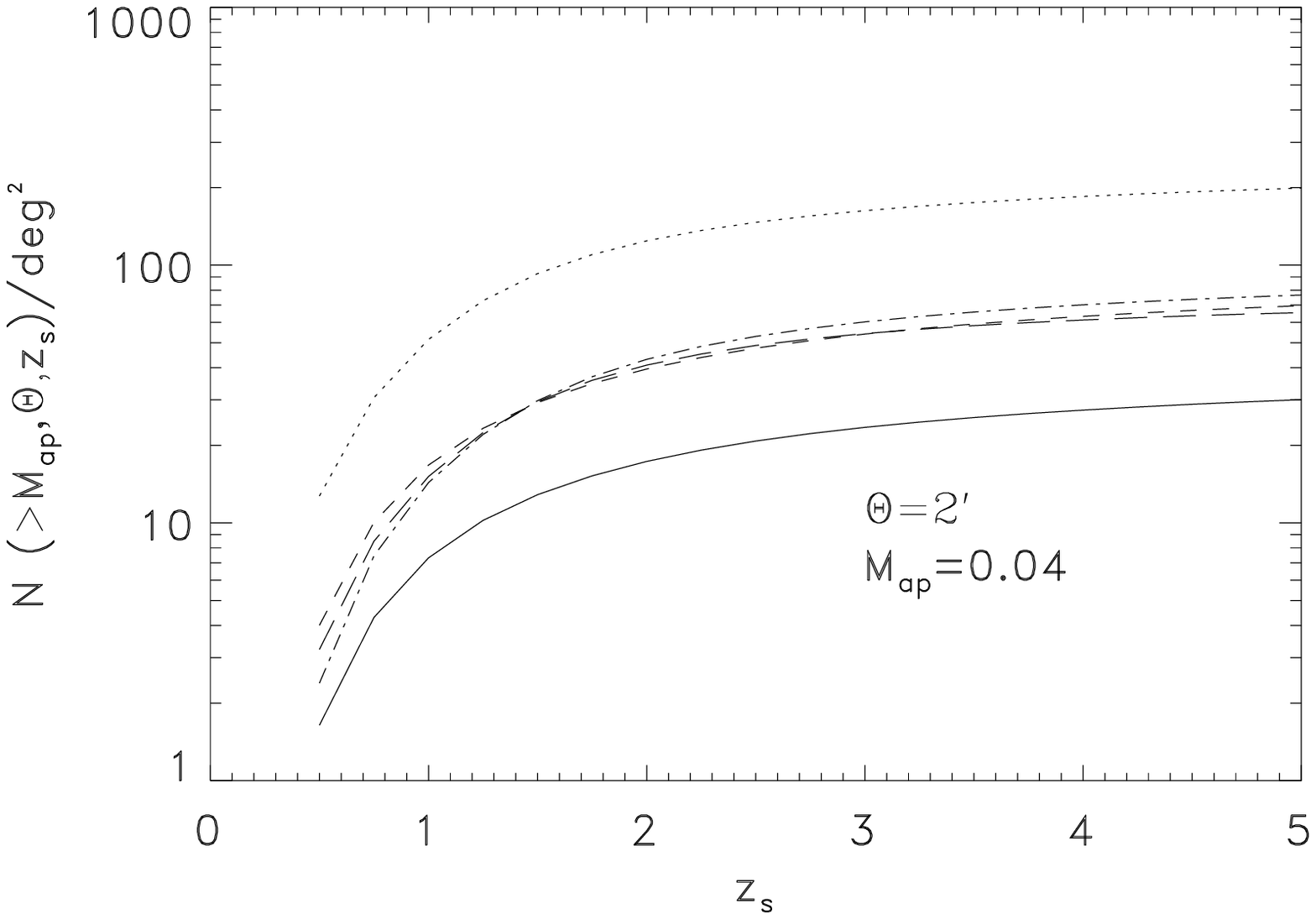}}
\caption{The number of haloes per square degree with aperture mass
greater than  $M_{\rm ap}=0.04$,
as defined in (\ref{number}), as a
function of source redshift for the same cosmological models as indicated
in Figure \ref{map}. All sources are assumed to be at the same
redshift. The filter scale is $\theta=2$ arcmin.
\label{sredshift}}
\end{figure}

As expected from the evolution of the cluster mass function, the
volume elements and the aperture mass $M_{\rm ap}$, we get different
number densities of haloes for various cosmological parameters.  We
have a strong evolution with redshift in the $\Omega_{\rm d}=1$ model
and much less in the low-$\Omega_{\rm d}$ models. Furthermore, the
number density of rich clusters at intermediate redshifts drops more
rapidly in a critical density universe.  The dependence of the volume
elements and the aperture mass $M_{\rm ap}$ on the angular diameter
distances enhances the difference between the cosmologies for high
redshifts, and causes a decreasing number of haloes towards very small
redshifts.

If we integrate (\ref{nd}) over lens redshift we obtain the observable
$N (>M_{\rm ap},\theta)$. We have plotted this observable in Figure
\ref{numbermap} as a function of the aperture mass. The dependence of
(\ref{number}) on $M_{\rm ap}$ and $\theta$ can be understood as
follows: Since the aperture mass $M_{\rm ap}$ is a monotonically
increasing function of the halo mass (for a fixed redshift and filter
scale; see Figure \ref{map}) we expect $N (>M_{\rm ap},\theta)$ to
decrease with increasing $M_{\rm ap}$. If we enlarge the filter radius
the values of $M_{\rm ap}$ become smaller and, because of the monotony
of $M_{\rm ap}$ in the halo mass, for fixed $M_{\rm ap}$ the
corresponding threshold mass $M_{\rm t}$ increases. Therefore the
number of haloes decreases with increasing filter size.  Because of
this behaviour of $N (>M_{\rm ap},\theta)$ we can select a filter
radius and a value for $M_{\rm ap}$ which allows us to count a
sufficient number of haloes used for finding a significant difference
between the various cosmologies.  In practice we have to determine a
signal-to-noise ratio threshold above which we can consider a
significant detection. We will use here mainly a threshold value of
$S_{\rm c}=5$.

In Figure \ref{snoise} we have plotted the number of haloes per square
degree with aperture masses yielding a signal-to-noise ratio above the
threshold value $S_{\rm c}=5$ for different filter scales.  According
to Figure \ref{snoise} we count in all cosmologies the maximum number
of haloes for $\theta=2 \ \mbox{arcmin}$.  We will use this `optimal'
filter scale for our calculations.  According to (\ref{sig1}) the
corresponding aperture mass is $M_{\rm ap}=0.04$ for the `optimal'
filter radius and the signal-to-noise ratio threshold.

In real observations, the derived value of $M_{\rm ap}$ will
differ from the true one due to several effects. First, the intrinsic
ellipticity distribution of the source galaxies causes noise in the
measurement of $M_{\rm ap}$ which is given by (\ref{sigma}). Second,
the number of source galaxies in the filter will have at least
Possonian noise. And third, halos are not isolated, but there will be
perturbing mass inhomogeneities along the line-of-sight to the halo.
Comparing the first two sources of errors, the first dominates
(see Schneider et al. (1998)),
and so we consider $\sigma_{\rm c}$ as the uncertainty with which we
can measure $M_{\rm ap}$. The third source of error cannot be modelled
analytically, but must be estimated through ray-tracing simulations
such as those carried out by Jain, Seljak \& White (1998).
Then, by taking into account only the noise coming from the intrinsic
ellipticity distribution of the source galaxies, we assume that the
deviation $\Delta M_{\rm ap}$ between the true value of $M_{\rm ap}$
and the measured one $\hat M_{\rm ap}$ is Gaussian,
\begin{equation}
p(\Delta M_{\rm ap}, \theta) = \frac{1}{\sqrt{2 \pi} 
\ \sigma_{\rm c} (\theta)}
\mbox{exp} \left(- \frac{\Delta M_{\rm ap}^2}
{2 \ \sigma_{\rm c}^2 (\theta)} \right).
\label{prob}
\end{equation} 
In Figure \ref{numbermapcon} we have plotted the convolution
$\hat N(> \hat M_{\rm ap}, \theta)$ of
$N(>M_{\rm ap}, \theta)$ with the distribution (\ref{prob})
for the filter scale $\theta=2'$,
\begin{equation}
\hat N(> \hat M_{\rm ap}, \theta) = 
\int d M_{\rm ap} \ N(>M_{\rm ap}, \theta) \ 
p(\hat M_{\rm ap} - M_{\rm ap}, \theta).
\label{numberhat}
\end{equation}
In comparison with the
non-convolved function (see Figure \ref{numbermap}) the values
of (\ref{numberhat}) are only slightly enhanced for the values
of $M_{\rm ap}$ we are interested in (e.g., $M_{\rm ap}=0.04,0.08$). 
Therefore, in the following discussion we shall
neglect the difference between the distributions of $M_{\rm ap}$ and
$\hat M_{\rm ap}$.

We shall now discuss whether from measuring the number density of
haloes above a given threshold $M_{\rm ap}$, one can distinguish
between the various cosmological models mentioned at the beginning of
this section. From Figs.\ 5 and 6, we see that the EdS models with
$\sigma_8=0.6$ and $\Gamma=0.25$ [hereafter EdS(0.6,0.25)], and with
$\sigma_8=1$ and $\Gamma=0.25$ [hereafter EdS(1,0.25)], have a
considerably lower and higher, respectively, number density of haloes
for given $M_{\rm ap}$ than the three other models. From the numbers
in Table 1, considering a value of $M_{\rm ap}=0.04$ or
signal-to-noise of 5, it is clear that these two cosmologies can be
distinguished significantly (by which we mean that the Poisson error
bars do not overlap) from the other three already with 1 ${\rm deg}^2$
of a deep imaging survey. To distinguish between the other three
cosmologies [EdS(0.6,0.5), OCDM, $\Lambda$CDM], a larger-area survey
is needed. Taking the projected MEGACAM survey with its expected 25
${\rm deg}^2$ as an example (Mellier et al.\ 1998), one sees from the
numbers in Table 1 that 
at $M_{\rm ap}=0.08$ (signal-to-noise of 10), this survey is more than
sufficient to allow a clear distinction of these three cosmologies.
We therefore conclude that the currently planned wide-field imaging
surveys will allow to separate between the most popular currently
discussed cosmological models.

In order to get a more precise handle on the values of the
cosmological parameters and/or the shape of the initial power
spectrum, more detailed information may be used. Assuming that the
haloes giving rise to measurements of $M_{\rm ap}$ are not completely
dark, but cluster-like (though possibly with a broad range of
mass-to-light ratios), one might be able to identify a measured halo
with a galaxy overdensity on the sky and/or in redshift, and thus
determine the redshift of the corresponding halo, using either
photometric redshift techniques or spectroscopy. In this case, the
redshift dependence of the halo distribution can be measured. As shown
in Figs.\ \ref{ndachx} and \ref{ndach1}, the redshift evolution of the
halo density as probed by $M_{\rm ap}$ is quite different in the
cosmologies considered here.
 
In Table \ref{table} we have also displayed the number of haloes per
square degree with aperture mass greater than $M_{\rm ap}=0.04$ and
$M_{\rm ap}=0.08$ for the filter scale $\theta=2'$ for the five
cosmological models, using two different redshift intervals,
$z\in[0.15,0.4]$ and $z\in[0.4,1]$.  By comparing the halo densities
in the different redshift intervals for the various cosmologies in
Figs.\ \ref{ndachx} and \ref{ndach1}, we expect the largest
differences between the cosmological models for $M_{\rm ap}=0.08$.
The reason for this is the stronger evolution for the rich cluster
mass function which corresponds to large values of the aperture mass
(see Figure \ref{map}). Whereas the EdS(0.6,0.25) and EdS(1,0.25)
models are again very different from the other three, the use of
redshift information greatly helps to distinguish the EdS(0.6,0.5)
model from the two low-density models. For the latter, a survey area
of less than 3 ${\rm deg}^2$ would be sufficient.

One might think of another way to obtain redshift information, namely
to use source galaxies at different redshifts (distinguished, say, by
photometric redshift estimates). To investigate this effect, we have
plotted in Figure \ref{sredshift} the dependence of the number of
haloes on the redshift of the sources for $M_{\rm ap}=0.04$ and
$\theta=2'$.  All sources are assumed to be at the same redshift
$z_{\rm s}$. Whereas the number density of haloes as measured with
$M_{\rm ap}$ depends strongly on the source redshift, this dependence
is quite similar in all cosmologies, except at rather low redshifts,
$z_{\rm s}\sim 0.6$. However, their number density is likely to be
fairly small, so that the differences seen in Fig.\ \ref{sredshift}
will be very difficult to measure. We therefore discard this indicator
at this point.

\section{Discussion and Conclusions}

In this paper we investigated the statistics of high signal-to-
noise peaks in the aperture mass map in various cosmological models.
We constructed the observable number of peaks in the aperture mass 
$N(> M_{\rm ap}, \theta)$ using
the Press-Schechter theory for evaluating the number density of haloes
and the universal density profile of NFW. The observable number
density of high signal-to-noise peaks in the aperture mass map -- or
in other words, the number density of mass-selected haloes -- is large
in all cosmological models considered here, and range from $\sim
10 \ {\rm deg}^{-2}$ for a cluster normalized EdS model to $\sim 70 \ {\rm
deg}^{-2}$ for a COBE-normalized EdS model, quoted for a
signal-to-noise ratio of 5. Even for a signal-to-noise ratio of 10,
the number density of detectable haloes is about one per square
degree. Hence, in future wide-field imaging surveys, such haloes will
easily be found, so that a mass-selected sample of `clusters' is
within reach. Given that the cluster abundance has been used
extensively as a cosmological probe, this mass-selected sample will be
extremely useful to related observations to theoretical predictions. 

We estimated that a few square degrees of a deep wide-field
imaging survey will be sufficient to distinguish between some of the
most popular cosmological parameter sets. In particular,
cluster-normalized low-density universes can be easily distinguished
from a cluster-normalized EdS model, which is mainly due to the fact
that in the latter, the number density of haloes at a redshift of
$z_{\rm d}\sim 0.3$, which is mainly probed by our technique, 
is predicted to be considerably lower than in the
open models. 

Whereas our estimates on the number density of detectable haloes are
based on several simplifying assumptions (e.g., that halo number
density can be obtained from the Press-Schechter theory, that the mass
density is spherical and follows an NFW profile, that halos are
isolated, etc.) and therefore probably not very accurate, the numbers
obtained should approximately reflect the true situation. In
particular, the relative abundance as a function of $M_{\rm ap}$ and
in dependence on cosmological parameters will be the same as
calculated here. For more quantitative estimates, ray-tracing
calculations in a model universe obtained from N-body simulations have
to be used. With results obtained from there, more sensitive
statistics for the determination of cosmological parameters can be
derived.

\section*{Acknowledgments}
We want to thank C.S. Frenk for providing us the routine for computing
the NFW profile parameter.
This work was supported by the ``Sonderforschungsbereich 375-95 f\"ur
Astro-Teilchenphysik" der Deutschen
Forschungsgemeinschaft.

\bsp

\label{lastpage}


\begin{thebibliography}{99}
\bibitem{br} Bardeen, J.M., Bond, J.R., Kaiser, N. \& Szalay, A.S.
             1986, ApJ 304, 15
\bibitem{ba} Bartelmann, M. 1996, A\&A 313, 697
\bibitem{be} Bartelmann, M., Huss, A., Colberg, J.M., Jenkins, A.
             \& Pearce, F.R. 1998, A\&A 330, 1 
\bibitem{bo} Borgani, S., Rosati, P., Tozzi, P. \& Norman, C. 1998, 
             preprint
\bibitem{br} Brainerd, T.G., Blandford, R.D. \& Smail, I. 1996,
             ApJ 466, 623
\bibitem{ef} Efstathiou, G. 1996, in: {\it Cosmology and large scale
structure}, Les Houches Session LX, R. Scheffer, J. Silk, M. Spiro \&
J. Zinn-Justin (eds.), North-Holland, p.133.
\bibitem{ek} Eke, V.R., Cole, S. \& Frenk, C.S. 1996, MNRAS 282, 263
\bibitem{fo} Fort, B., Prieur, J.L., Mathez, G., Mellier, Y. \&
             Soucail,G. 1988, A\&A 200, L17
\bibitem{fo1} Fort, B. \& Mellier, Y. 1994, A\&AR5, 239
\bibitem{ja} Jain, B., Seljak, U. \& White, S.D.M. 1998, astro-ph/
              9804238
\bibitem{ka} Kaiser, N. \& Squires, G. 1993, ApJ 404, 441
\bibitem{ka1} Kaiser, N., Squires, G., Fahlman, G. \& Woods, D. 1994,
              astro-ph/9407004
\bibitem{ka2} Kaiser, N. 1995, ApJ 1995, L1
\bibitem{la} Lacey, C. \& Cole, S. 1993, MNRAS 262, 627
\bibitem{lu} Luppino, G.A. \& Kaiser, N. 1997, ApJ 475, 20
\bibitem{me} Mellier, Y., van Waerbeke, L. \& Bernardeau, F. 1998, preprint
\bibitem{na} Navarro, J.F, Frenk, C.S. \& White, S.D.M. 1996, 
             ApJ 462, 563
\bibitem{na} Navarro, J.F, Frenk, C.S. \& White, S.D.M. 1997, 
             ApJ 490, 493
\bibitem{pr} Press, W.H. \& Schechter, P. 1974, ApJ 187, 425
\bibitem{sc} Schneider, P. 1996, MNRAS 283, 837
\bibitem{sc1} Schneider, P., van Waerbeke, L., Jain, B. \& Kruse, G.
              1998, MNRAS 296, 873
\bibitem{se} Seitz, S., Saglia, R.P., Bender, R., Hopp, U., Belloni,
             P. \& Ziegler, B., astro-ph/9706023
\bibitem{so} Soucail, G., Fort, B., Mellier, Y. \& Picat, J.P.
             1987, A\&A 184, L14
\bibitem{sq} Squires, G. \& Kaiser, N. 1996, ApJ 473, 65
\bibitem{ty} Tyson, J.A., Valdes, F. \& Wenk, R.A. 1990, ApJ 349, L1 
\bibitem{we} Webster, R.L. 1985, MNRAS 213, 871
\bibitem{wh} White, S.D.M., Efstathiou, G. \& Frenk, C.S. 1993, MNRAS
             262, 1023
\bibitem{ye} Yee, H.K.C., Ellingson, E., Bechtold, J., Carlberg, R.G.
             \& Cuillandre, J.-C. 1996, AJ 111, 1783

\end{thebibliography}
\end{document}